# Characterization of microwave absorption in carbon nanotubes using resonance aperture transmission method

O. Malyuskin,[1,a)] P.Brunet,[2] D. Mariotti,[2] R. McGlynn,[2] P. Maguire,[2]

[1]ECIT, Queen's University Belfast, Belfast, BT3 9DT, UK, [2]NIBEC, Ulster University, BT37 0QB, Newtownabbey, UK
a) The author to whom correspondence should be addressed o.malyuskin@qub.ac.uk

**Abstract**: A new method to characterize microwave electromagnetic (EM) absorption of a bulk carbon nanotube (CNT) material is proposed and experimentally evaluated in this paper. The method is based on the measurement of microwave transmission through a capacitive-resonator aperture (CRA) in a conductive screen loaded with a CNT sample under test. This method allows to measure microwave permittivity and absorption of thin samples (~ 0.1μm-10μm thick) with linear dimensions much smaller than the wavelength of radiation in free space. This "minimal" sample requirement restricts the application of conventional microwave characterization methods such as free-space or waveguide permittivity characterization. It is demonstrated that the resonance E-field enhancement inside the CRA leads to strong EM interaction of the microwave E-field with the CNT sample under test thus enabling high sensitivity and dynamic range (~ 5dB) of the measurement procedure. Another advantage of the proposed technique over conventional non-resonance characterization methods is that in the resonance transmission band, the CRA operation is reflection-less which leads to a relatively simple qualitative algebraic de-embedding procedure of the material parameters based on the principle of energy conservation. The experimental microwave absorption data of the multiwall CNT samples are presented in the S frequency band (2-4GHz), demonstrating microwave absorption properties of the multiwall CNT ribbons.

## I. INTRODUCTION

Development of advanced composite and nano-materials requires a range of measurement techniques to characterize the electrical, electronic, mechanical and thermal properties of the material. One important class of materials is represented by thin films and self-standing nano-materials with sample thickness varying from mono-atomic layers to few micrometers with relatively large, macroscopic area dimensions. An example of such a material is CNT ribbons which exhibit very unique mechanical, electric and electronic properties[1].

CNTs have a wide range of unique applications in nano- and microelectronics, radiation detection, chemical and bio-sensing, composite materials, etc[2,3]. In this letter, a new method to characterize EM properties of a bulk CNT material in the microwave range is proposed. The method is based on the reflection-less resonance transmission of microwave radiation through the CRAs with and without a CNT sample. From the measurement data, real and imaginary parts of the CNT complex permittivity can be extracted using differential material parameters de-embedding[4,5]. Permittivity of the pure CNT can be further employed to design EM absorbers and EM interference shielding based on the composite materials formed by the CNT material inclusions inside a host medium, e.g. epoxy or resins[6,7]. These composite materials can be promising for advanced aerospace applications[3,6-8] and future wireless communications, especially, for the design of the intelligent EM-controlled metasurfaces[8], due to mechanical and chemical stability and very low weight density of the CNTs.

The existing EM material characterization methods[10-14] can be divided into several groups including free-space, waveguide, resonating cavity, 3D resonator, transmission line, open coaxial, evanescent probe and a parallel-plate capacitor method. The discussion of the application range, advantages and drawbacks of these methods can be found elsewhere[10-14]. As a rule, most of the abovementioned methods require sufficiently large CNT samples covering the area of at least $0.25\lambda \times 0.5\lambda$, where $\lambda$ is a wavelength of radiation in free space. Also, the measurement of thin samples (thickness of few micrometers or less) represents significant challenge for free-space or waveguide methods in microwave range ($\lambda$ is in the range of 0.1cm-30cm) due to low sensitivity and numerical errors in permittivity de-embedding algorithm in the thin sample permittivity reconstruction scenario[14].

Resonance CRA transmission method discussed here offers a number of distinctive advantages: high sensitivity and dynamic range of the method is due to EM field enhancement inside the CRA leading to strong EM interaction between the testing microwave and even very thin samples. Electrically small size of the CRA allows to use "minimal" thin CNT samples with characteristic 2D dimensions significantly smaller than the radiation wavelength.

## II. GEOMETRY OF THE MEASUREMENT SETUP

The measurement setup geometry is shown in Fig. 1. The setup consists of a pair of dipole antennas and a CRA in the conductive screen loaded with a thin CNT material sample. The transmit (TX) and receive (RX) microwave signals are generated and received by the vector network analyzer (VNA).

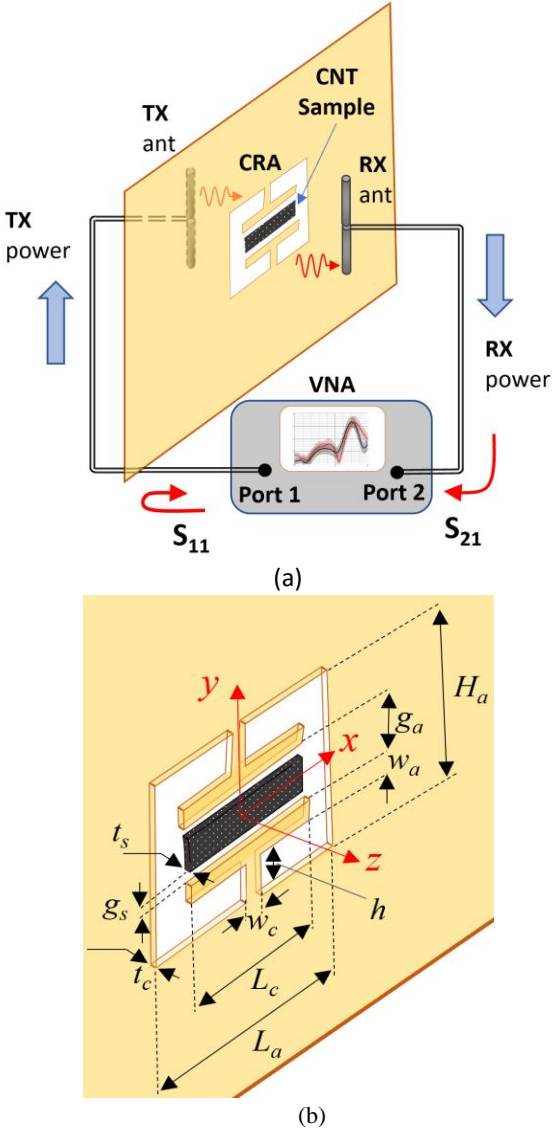

FIG. 1. (a) Measurement setup geometry; (b) detailed view of the CNT-loaded aperture with dimensions. The measurement setup consists of a CRA, a pair of transmit (TX) and receive (RX) dipole antennas and a vector network analyzer (VNA) to measure the microwave S-parameters.

This material characterization method is based on differential measurements, when the transmission data through the loaded and unloaded aperture are required along with the measurement of the free-space transmission link between the RX and TX antennas, in order to de-embed antenna characteristics and free-space propagation link properties from the overall measurement system transmission.

## III. ELECTROMAGNETIC MODEL OF THE CRA MICROWAVE TRANSMISSION

Microwave transmission through the CRA aperture can be described[10, 12, 15] by the system of equations (1)-(2):

$$\boldsymbol{E}_{t_{FS}}(\boldsymbol{r}) = \hat{T}(\varepsilon_0) \cdot \boldsymbol{E}_i(\boldsymbol{r}_A), \quad (1)$$

$$\boldsymbol{E}_{t_{CNT}}(\boldsymbol{r}) = \hat{T}(\varepsilon_{CNT}) \cdot \boldsymbol{E}_i(\boldsymbol{r}_A), \quad (2)$$

where $\boldsymbol{E}_{t_{FS}}$ - is a TX microwave *E*-field through the unloaded (in the air) CRA, $\boldsymbol{E}_{t_{CNT}}(\boldsymbol{r})$ is a TX *E*-field through the CRA loaded with a CNT sample, $\boldsymbol{r}$ is a radius-vector of observation point in the z>0 half-space, $\boldsymbol{r}_A$ is a radius-vector of observation point across the CRA aperture, in the plane z=0, Fig.1(b). $\hat{T}$ is a second-rank (3x3) tensor operator describing microwave TX properties of the CRA. $\boldsymbol{E}_i$ is the incident EM field generated by a TX antenna with the center located at z=-$z_d$. To derive the transmission equations in terms of the measurable RX power, the TX antenna is characterized by its radiation gain $\boldsymbol{G}_{TX}(\boldsymbol{r}, \boldsymbol{r}_{TX})$, where $\boldsymbol{r}_{TX}$ is a radius-vector of the TX antenna center. The dipole RX antenna can be characterized by its effective vector length, $\boldsymbol{l}_{RX}$, relating RX voltage across the RX antenna terminal and the E-field acting on the antenna. Assuming that the input impedances of the RX and TX antennas are the same (50 ohm), it is straightforward to derive the transmission equation in terms of the TX and RX power and EM transmission tensor $\hat{T}$ of the CRA,

$$P_{RX} = \left(\hat{T} \cdot \boldsymbol{G}_{TX} \cdot \boldsymbol{l}_{RX}\right)^2 P_{TX} \quad (3)$$

Differential measurement procedure is based on the normalization of the transmitted power through the CNT-loaded CRA, $P_{RX,CNT}$, by the transmitted power through the unloaded CRA, $P_{RX,FS}$ for the same input power $P_{TX}$

$$P_{RX,CNT}/P_{RX,FS} = [T(\varepsilon_{CNT})/T(\varepsilon_0)]^2 \quad (4)$$

where $T(\varepsilon_{CNT})$ and $T(\varepsilon_0)$ are the EM transfer functions of the loaded and unloaded CRA for the vertically y-polarized *E*-field. Differential measurement method allows to eliminate RX and TX antenna characteristics from the CNT permittivity de-embedding procedure which employs normalized transmitted power function (4).

The microwave transmission function of the CRA appeared in (1), (2) is an integral-differential tensor operator, whose form can be established using the EM field potential equations. For the monochromatic, $exp(-i\omega t)$, time dependence the EM field transmitted through the material-loaded aperture can be written in the form[16, 17]

$$E_t(r) = \frac{i}{\omega\mu\varepsilon_n}(\nabla\nabla + k_n^2) \cdot A - \frac{1}{\varepsilon_n}\nabla \times F \quad (5)$$

In (5) µ is a magnetic permeability of free space, index $n$ defines the material, $n$ = CNT or $n$ = free space or air, $k_n = \sqrt{\varepsilon_n}\,\omega/c$, $c$ is a speed of light in vacuum, and $A$ and $F$ are the EM vector-potentials[17],

$$A(r) = \mu \int_{Apert.} G_n(r, r') J(r') dr', \quad (6)$$

$$F(r) = \varepsilon_n \int_{Apert.} G_n(r, r') J_m(r') dr' \quad (7)$$

In (6) the integration is carried out over the volume occupied by the CNT material, Fig.1(b), and involves the electric polarization current density $J$

$$J(r) = -i\omega\varepsilon_0(\varepsilon_{rCNT} - 1)E_a(r), \quad (8)$$

where $\varepsilon_{rCNT} = \varepsilon'_{CNT} + i\varepsilon''_{CNT}$ is a complex-valued relative permittivity of CNT, $E_a(r)$ is an E-field inside the CNT volume. In (7), the area integration is carried out inside the CRA area $-L_a/2 < x < L_a/2$, $-H_a/2 < y < H_a/2$, free from the CNT or conductor, at the plane $z = t_c$, and involves magnetic current density $J_m$

$$J_m(r) = -z_0 \times E_a(r) \quad (9)$$

Finally, the scalar Green's function in (6), (7) is given by a relation

$$G(r, r') = exp(ik_n|r - r'|)/4\pi|r - r'| \quad (10)$$

The explicit form of the transfer function $T$ in (1), (2) can be established by using permittivity $\varepsilon_{CNT}$ in (5) – (10) and calculating the E-field $E_a(r_A)$ inside the CRA from the boundary conditions and equations (6), (7). It should also be noted that the real $\varepsilon'_{CNT}$ and imaginary $\varepsilon''_{CNT}$ parts of the CNT permittivity $\varepsilon_{CNT}$ appear in the transmission equation (4) in a coupled form, thus permittivity de-embedding requires the inverse problem solution of (1)-(10) based on the full-wave numerical computation. Below, an analytical solution based on the energy conservation principle is proposed which can also serve as an initial approximation in the full-wave numerical solution to (1)-(10).

## IV. ANALYTICAL SOLUTION FOR THE CNT PERMITTIVITY DE-EMBEDDING

To obtain an approximate analytical solution to the permittivity of the CNT sample under test, unloaded and loaded CRAs can be represented as equivalent circuits[18], Fig.2.

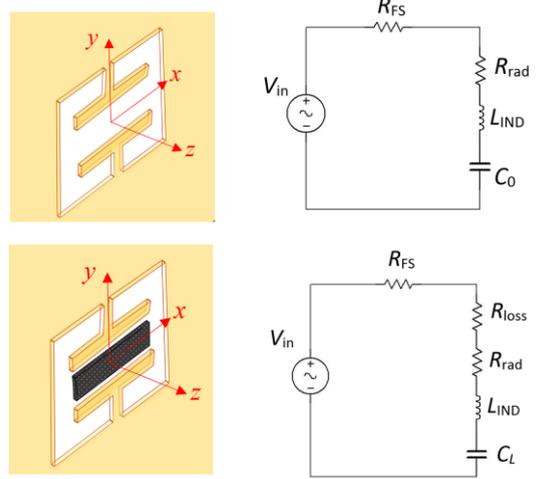

FIG. 2. Equivalent lump circuit model of an unloaded (top) and CNT-loaded (bottom) CRA.

In these circuits, source $V_{in}$ models TX antenna, $R_{FS}$ stands for the free-space characteristic impedance, $R_{rad}$ is CRA radiation resistance, $L_{IND}$ is a CRA equivalent inductance and $C_0$ and $C_L$ is unloaded and loaded equivalent capacitance. $R_{loss}$ models the microwave loss due to conductivity (imaginary part of the permittivity) of the CNT sample, while loaded capacitance $C_L$ is a real-valued function of CRA geometry and real part of the CNT permittivity. Since the CNT material studied in this work is a lossy non-magnetic or weakly-magnetic dielectric, equivalent inductance of the loaded and unloaded aperture is approximately the same. The exact form of the equivalent circuit lump parameters can be established using surface currents and charge density[16,17] across the CRA, however this is not required for the analytical model described below.

Using resonance circuit models in Fig.2, it can be seen that CRA loading with a CNT sample results in the microwave transmission resonance frequency shift,

$$f_0^2/f_L^2 = C_L/C_0 \quad (11)$$

where $f_0$ and $f_L$ are the resonance frequencies of the unloaded and loaded CRA respectively. Due to the high-quality resonance of the CRAs used in this work

(measured Q-factor typically exceeds 10), resonance frequency pulling[19], $f_L^{loss}$,

$$f_L^{loss} = f_0\sqrt{1 - 1/4Q^2}$$

due to microwave loss caused by the CNT absorption is small (less than 5%), and therefore the real part of CNT permittivity plays a dominant role in the CRA transmission resonance frequency change. To derive analytical expressions for the resonance frequency shift as a function of the CNT permittivity real part, CNT-loaded and unloaded capacitance of a CRA can be evaluated using a quasi-static approximation (please refer to Appendix A for details)

$$(\varepsilon'_{rCNT} + A_0)/(1 + A_0) = f_0^2/f_L^2 \quad (12)$$

where the constant $A_0 = \pi g_a/(w_a \ln(1 + g_a/w_a))$ depends on the CRA capacitive resonator gap to width of the plate's ratio. Formula (12) is obtained in the assumption that the CNT material fills uniformly the gap space $-L_c/2 \leq x \leq L_c/2$, $-\frac{g_a}{2} \leq y \leq \frac{g_a}{2}$, $0 \leq z \leq t_c$ between the CRA plates 1 and 2, FIG.5, Appendix A.

The imaginary part of the CNT sample permittivity can be found from the power conservation principle, which results in the equation (13), detailed derivation is provided in Appendix B,

$$\varepsilon''_{rCNT} = \varepsilon'_{rCNT}[W_1 P_{RX}(f_0) - W_2 P_{RX}(f_L)] \quad (13)$$

In equation (13)

$$W_1 = \frac{D(f_L)}{D^2(f_0)}\frac{T(f_L)}{T(f_0)}\frac{1+Q_L}{1+Q_0}R(f_0), W_2 = \frac{R(f_L)}{D(f_L)}, \quad (14)$$

Equations (13), (14) involve measurable parameters: $D(f_0)$ and $D(f_L)$ are the propagation path loss factors between the TX antenna and the CRA plane at resonance frequencies $f_0$ and $f_L$ respectively, coefficients $T$ describe TX antenna matching, $T(f) = |1 - S_{11}(f)|$, coefficients $R$ characterize matching of the RX antenna at frequency $f$, $R(f) = |1 - S_{22}(f)|^{-1}$, Q-factors of the loaded ($Q_L$) and unloaded ($Q_0$) CRAs can be calculated from the CRA transmission power characteristics in standard form as a ratio of the resonance frequency to the 3dB bandwidth, $Q = f_{res}/\Delta f$, $P_{RX}(f)$ is RX power measured by the RX antenna at frequency $f$, normalized to the input power at TX antenna. Equations (12), (13) allow to de-embed the real $\varepsilon'_{rCNT}$ and imaginary $\varepsilon''_{rCNT}$ parts of the CNT relative permittivity $\varepsilon_{rCNT}$ based on algebraic equations.

## V. EXPERIMENTAL RESULTS

### A. Sample preparation

CNTs were made in an axial tubular high temperature furnace at 1290 °C known as a floating catalyst chemical vapor deposition (FC-CVD) in background hydrogen (1350 sccm). The precursors used for the CNT production, are ferrocene for the catalyst's nanoparticles, thiophene to supply sulfur as a growth limiter of the CNT diameter and methane as a carbon precursor. Both ferrocene and thiophene are introduced via bubblers with hydrogen flow at 130 sccm and 90 sccm, respectively. The bubbler with thiophene was kept at 25°C. The decomposition of the ferrocene precursor leads to nucleation and growth of iron nanoparticles with incorporation of sulfur, which has been shown to play an important role in these mechanisms[20,21]. Methane flow was set at 160 sccm. During the synthesis process, an aerogel forms at the exit of the chamber and this is collected and pulled out from the furnace by the help of a rod. This process allows the formation of macroscopic assemblies of multiwall CNTs that can be adapted in different morphologies.

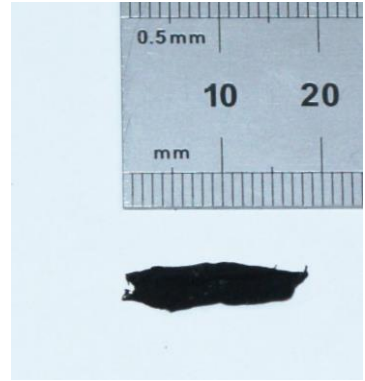

(a)

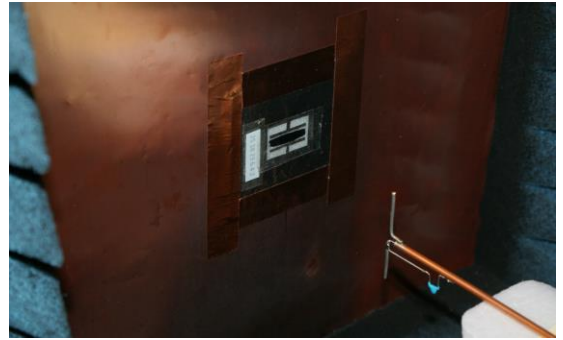

(b)

FIG. 3. (a) Example of a CNT sample; (b) CNT-loaded CRA and TX dipole antenna inside anechoic environment

In this case, CNTs have been mechanically compacted and cut in length to produce CNT ribbons of a few centimeters and a thickness with the range of 1-4 μm.

The CRAs have been fabricated using thin aluminum foil of approximately 10 μm thickness. The CNT samples were positioned on the Rohacell 5mm-thick substrate IG51 (permittivity 1.071, tan δ =0.0031) and affixed with Scotch tape (permittivity 1.9, tan δ =0.006). Four CRAs were fabricated with the dimensions summarized in Table I. In all CRAs $w_s=w_a$, $<g_s>$ is an average value of gap $g_s$.

### B. Microwave measurement data and analysis

Two sets of dipole antennas were used to provide optimal matching in the bands 1.7-2.5GHz and 2.4-3.2GHz. The return loss $|S_{11}|$ characteristics of the antennas in measurement setup are provided in Appendix C. This choice of the frequency range is chosen due to practical interest in the WiFi band CNT characteristics and ease of the measurements at lower frequency range. However, the design can be straightforwardly scaled for higher frequency range.

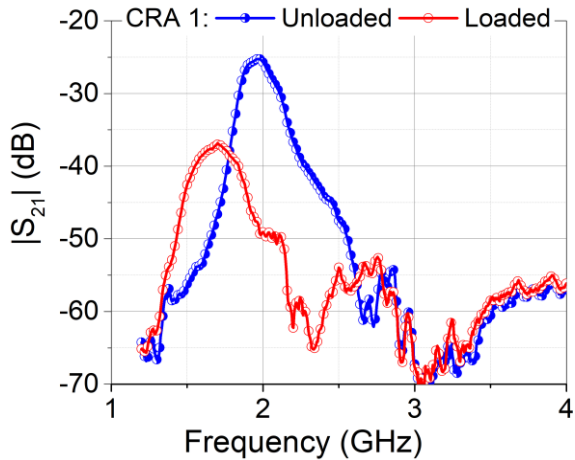

(a)

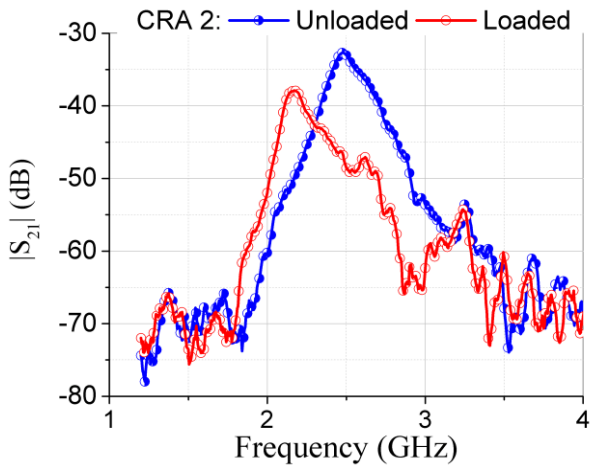

(b)

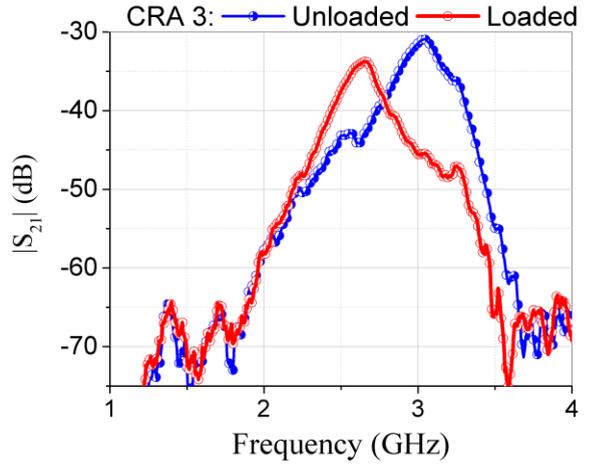

(c)

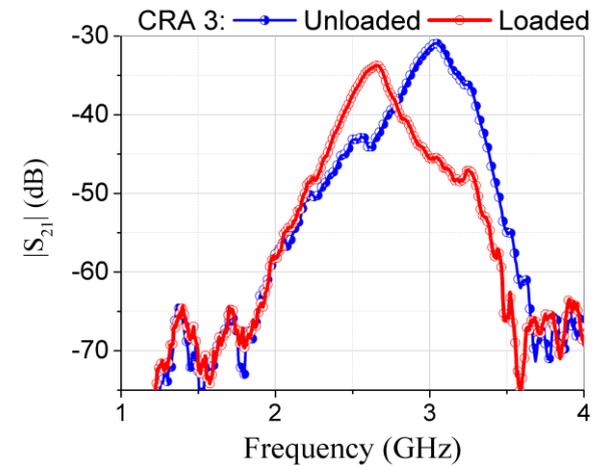

(d)

FIG. 4. (a)-(d) Measured transmission power functions $|S_{21}|$ of four fabricated CRAs, unloaded and loaded with a CNT sample.

The CRA transmission characteristics $|S_{21}|$ are presented in FIG.4. Table II summarizes the measured parameters: resonance frequency, $Q$-factors, RX power and de-embedded CNT permittivity using an approximate model (13)-(14). The dipole antennas are fed with -10dBm power (at the VNA output).

FIG.4. demonstrate very high dynamic range and sensitivity of the CRA to the CNT sample loading, the resonance frequency shift in all cases is around 11% - 14%, RX power variation between unloaded and loaded CRA $|S_{21}|$ is approximately 2-10dB which is due to field enhancement inside the CRA (please see more detail in the Appendix D). The permittivity data calculated in Table II are consistent with the previous measurements carried out by other authors[22-26], it is believed that the major absorption mechanism in the studied CNT ribbons is due to CNT conductivity losses[26].

TABLE I. Dimensions of the CRA (mm).

| No. CRA | $L_a$ | $H_a$ | $L_c$ | $g_a$ | $w_a$ | $<g_s>$ |
|---|---|---|---|---|---|---|
| 1 | 30 | 22 | 28 | 6 | 2 | 0.2 |
| 2 | 26 | 18 | 24 | 6 | 2 | 0.3 |
| 3 | 22 | 16 | 20 | 6 | 2 | 0.5 |
| 4 | 20 | 15 | 18 | 6 | 2 | 0.8 |

TABLE II. Measured resonance frequency, Q-factors, RX power and de-embedded permittivity of CNT

| No. CRA | $f_0$ (GHz) | $f_L$ (GHz) | $Q_0$ | $Q_L$ | $P_{RX}(f_0)$ (dB) | $P_{RX}(f_L)$ (dB) | $Re\ \varepsilon_{CNT}$ | $Im\ \varepsilon_{CNT}$ |
|---|---|---|---|---|---|---|---|---|
| 1 | 1.97 | 1.71 | 9.34 | 5.89 | -25.25 | -37.02 | 3.55 | 3.97 |
| 2 | 2.48 | 2.175 | 13.41 | 14.51 | -32.71 | -37.82 | 3.34 | 2.09 |
| 3 | 3.04 | 2.66 | 13.20 | 12.98 | -30.90 | -33.76 | 3.35 | 2.69 |
| 4 | 3.27 | 2.94 | 16.86 | 13.386 | -32.82 | -33.10 | 2.85 | 1.86 |

It should be noted that additional measurements not presented here for the brevity sake demonstrate the CRA transmission sensitivity to the gap $g_s$ between the CNT sample and the CRA plates. This gap leads to the lower values of permittivity as can be understood from the effective medium approximation[27], therefore the gap $g_s$ should be as minimal as possible for accurate permittivity characterization. The major drawback of the proposed method in its present form is a requirement for broadband dipole antennas matched in the wide frequency band. This drawback can be eliminated by using printed resonators fed by broadband printed strip lines and incorporating printed array of the CRA resonators on a single printed circuit board.

## VI. CONCLUSIONS

A new method to characterize permittivity of the CNT samples using resonance, reflection-less CRA transmission is proposed. Initial experimental data demonstrate high sensitivity and dynamic range of the proposed method, which is practically important for the EM characterization of thin "minimal" samples of the CNT material. Approximate analytical procedure of the permittivity de-embedding from the measurement data is proposed, reconstructed permittivity values are consistent with the previous measurements. This method can be easily scaled in the frequency range (from microwave to THz) thus providing an accurate EM characterization tool in addition to existing methods.

## REFERENCES

[1]J. Wang, X. Luo, T. Wu, et al., "High-strength carbon nanotube fibre-like ribbon with high ductility and high electrical conductivity". Nat Commun 5, 3848 (2014)

[2]*Carbon nanotubes. Science and applications*. Ed. M. Meyyappan, CRC Press 2005.

[3]*Advanced nanomaterials for aerospace applications*. Ed. C. Cabrera, F. Miranda, Pan Stanford Publishing 2014.

[4]K. Y. You, F. B. Esa and Z. Abbas, "Macroscopic characterization of materials using microwave measurement methods — A survey," PIERS – FALL 194-204 (2017) doi: 10.1109/PIERS-FALL.2017.8293135

[5]F. Costa, M. Borgese, M. Degiorgi, A. Monorchio, "Electromagnetic Characterisation of Materials by Using Transmission/Reflection (T/R) Devices," *Electronics* 6, 95 (2017). doi:10.3390/electronics6040095

[6]H. Zhang, G. Zeng, Y. Ge, T. Chen and L. Hu, "Electromagnetic Characteristic and Microwave Absorption Properties of Carbon Nanotubes/Epoxy Composites in the Frequency Range from 2 to 6 GHz," *Journ. Applied Physics*, Vol. 105, No. 5, 2009, Article ID: 054314.

[7]X. Xia, Y. Wang, Z. Zhong, George J. Weng, "A theory of electrical conductivity, dielectric constant, and electromagnetic interference shielding for lightweight

graphene composite foams," *Journ. Appl. Phys*. 120, 085102 (2016).

[8]L. Wang, Zhi-Min Dang, "Carbon nanotube composites with high dielectric constant at low percolation threshold," *Appl. Phys. Lett*. 87, 042903 (2005); https://doi.org/10.1063/1.1996842

[9]A. Li, Z. Luo, H. Wakatsuchi, et al., "Nonlinear, active, and tunable metasurfaces for advanced electromagnetics applications," *IEEE Access* 5, 27439–27452 (2017).

[10]F. Gonçalves, A. Pinto, R. Mesquita et al., "Free-Space Materials Characterization by Reflection and Transmission Measurements using Frequency by-Frequency and Multi-Frequency Algorithms," *Electronics* 7, 260 (2018). doi:10.3390/electronics7100260

[11]Y. Gao, M. T. Ghasr, M. Nacy, R. Zoughi, "Towards Accurate and Wideband In Vivo Measurement of Skin Dielectric Properties," *IEEE Trans. Instr. Meas*. 68(2) 512-524 (2019).

[12]J. Baker-Jarvis, M. D. Janezic, D. C. Degroot, "High-Frequency Dielectric Measurements: Part 24 in a Series of Tutorials on Instrumentation and Measurement," *IEEE Instr. Meas. Magazine* 13(2):24-31 (2010).

[13]L. Liu, et al., *Microwave Dielectric Properties of Carbon Nanotube Composites*. In *Carbon Nanotubes* (Intech Open 2010). doi: 10.5772/39420

[14]L. Liu, L. Kong, W.-Y. Yin, S. Matitsine, "Characterization of Single- and Multi-walled Carbon Nanotube Composites for Electromagnetic Shielding and Tunable Applications," *IEEE Trans. Electromagn. Compat*., 53, 943–949 (2011).

[15]J. Sheen, "Comparisons of microwave dielectric property measurements by transmission/reflection techniques and resonance techniques", *Meas. Sci. Tech*., 20, 042001 (2009).

[16]C. L. Gardner, G.I. Costache, "The Penetration of EM Waves Through Loaded Apertures," *IEEE Trans. Electromagn. Compat*. 3, 37, 358-366 (1995).

[17] J.A. Kong, *Electromagnetic Wave Theory*, Wiley-Blackwell (2000).

[18]A. W. Love, "Equivalent circuit for aperture antennas," in Electronics Letters, 13, 23, 708-710, 18 (1987).

[19]*Dielectric resonators*, Ed. D. Kajfez, P.Guillon, Artech House (1986).

[20]C. Hoecker, et al., "The Dependence of CNT Aerogel Synthesis on Sulfur-driven Catalyst Nucleation Processes and a Critical Catalyst Particle Mass Concentration," *Scientific Reports*, 7(1), 14519 (2017).

[21] M. Motta, et al., "The Role of Sulphur in the Synthesis of Carbon Nanotubes by Chemical Vapour Deposition at High Temperatures", *Journal Nanoscience Nanotechnology*, 8, 2442-9 (2008).

[22]D. Micheli, R. Pastore, C. Apollo, et.al., "Broadband electromagnetic absorbers using carbon nanostructure-based composites. *IEEE Trans. Microw. Theory Tech*., 59, 2633–2646 (2011).

[23]P. Savi, M. Giorcelli, S. Quaranta, "Multi-Walled Carbon Nanotubes Composites for Microwave Absorbing Applications," *Appl. Sci*. 9, 851 (2019).

[24]Z. Wang, G. Zhao, "Microwave Absorption Properties of Carbon Nanotubes-Epoxy Composites in a Frequency Range of 2 - 20 GHz," *Open Journal of Composite Materials*, 3 (2),17-23 (2013).

[25]A. Katsounaros, K. Z. Rajab, Y. Hao, et al. "Microwave characterization of vertically aligned multiwalled carbon nanotube arrays," *Appl. Phys. Lett*. 98, 203105 (2011).

[26]M. Green, X. Chen, "Recent progress of nanomaterials for microwave absorption," *Journal Materiomics*, 5(4), 503-541(2019).

[27]F. Hu, J. Song and T. Kamgaing, "Modeling of multilayered media using effective medium theory," 19th Meeting Elect. Performance of Electronic Packaging and Systems, 225-228 (2010). doi: 10.1109/EPEPS.2010.5642584.

[28]F. R. Zypman, "Mathematical expression for the capacitance of coplanar strips", *Journ. Electrostatics* 101 103371, (2019).

## APPENDIX A. CRA CAPACITANCE CALCULATION

Quasi-static capacitance of the unloaded and loaded CRAs can be evaluated considering the geometry of the CRA resonator, Fig.S1.

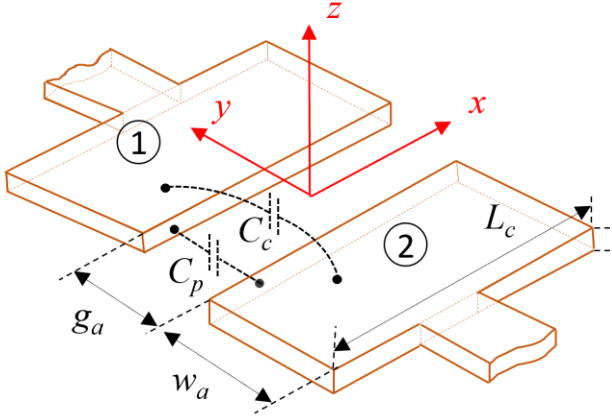

FIG. 5. Capacitance between the capacitive resonator conductive faces of the CRA

The quasi-static part of the capacitance between the CRA plates 1 and 2 corresponds to the first term in the $E$-field representation through the scalar potential[16],

$$E(r) = -\frac{1}{4\pi\varepsilon}\nabla\left[\int\frac{\rho(r')dr'}{|r-r'|} + \int\frac{\exp ik|r-r'|-1}{|r-r'|}\rho(r')dr'\right] \quad (15)$$

where $\rho(r')$ is a charge surface density across the surface of 1 and 2, and the field is calculated inside the gap $g_a$. Neglecting other terms except for the quasi-static part of the CRA equivalent capacitance is an accurate approximation when $k|r-r'| \leq 1/4$, i.e. when the CRA is electrically small. Since the charge and potential distribution across the surface of the CRA plates is non-uniform in the high-frequency case, average capacitance per unit area needs to be used in derivation of (12). The total CRA capacitance per unit area is a sum of capacitance per unit area between the parallel faces $Cp$ and capacitance per unit area between the coplanar faces $Cc$, Fig. 5 Taking this into account, the $C_L/C_0$ can be written as follows

$$\frac{C_L}{C_0} = \frac{C_p^L/t_c L_c + C_c^L/L_c w_a}{C_p^0/t_c L_c + C_c^0/L_c w_a} \quad (16)$$

where the capacitance between the parallel faces is given by $C_p^L = \varepsilon_0 \varepsilon_{rCNT} t_c L_c / g_a$, $C_p^0 = \varepsilon_0 t_c L_c / g_a$, and capacitance between the coplanar faces[28] is $C_c^L = C_c^0 = \pi\varepsilon_0 L_C / \ln(1 + g_a/w_a)$. Equation (12) follows directly from (16).

## APPENDIX B. POWER BALANCE EQUATION

The imaginary part of the CNT sample defines power dissipation; therefore, the power conservation principle can be employed to derive an equation for the imaginary part of the permittivity. The balance of microwave power delivered to the CRA from the TX antenna and radiated (transmitted), stored (in the reactive near field) and dissipated due to CNT material loss can be written as

$$D(f_0)T(f_0)P_{in} = P_{rad}(f_0) + P_{st}(f_0) \quad (17)$$

$$D(f_L)T(f_L)P_{in} = P_{rad}(f_L) + P_{st}(f_L) + P_{loss}(f_L) \quad (18)$$

Equations (17), (18) describe the power balance for the unloaded and CNT-loaded CRAs, respectively, $D(f_0)$ and $D(f_L)$ are the propagation path loss factors between the TX antenna and the CRA plane at resonance frequencies $f_0$ and $f_L$. These factors can be found experimentally by measuring the $S_{21}$ parameter at several distances between the RX and TX antennas. Coefficients $T$ describe TX antenna matching, $T(f) = |1 - S_{11}(f)|$. $P_{rad}(f)$ is a radiated power through the CRA which is related to the RX power measured by the RX antenna as follows

$$D(f)P_{rad}(f) = R(f)P_{RX}(f) \quad (19)$$

In (19), coefficient $R$ characterizes matching of the RX antenna at frequency $f$, $R(f) = |1 - S_{22}(f)|^{-1}$. The stored power $P_{st}$ can be expressed through the Q-factors of the loaded ($Q_L$) and unloaded ($Q_0$) CRAs, radiated power and dissipated power $P_{loss}$ due to CNT microwave absorption [18],

$$P_{st}(f_0) = Q_0 P_{rad}(f_0),$$

$$P_{st}(f_0) = Q_L[P_{rad}(f_0) + P_{loss}(f_L)] \quad (20)$$

Taking into account that $P_{loss} = \varepsilon''_{rCNT}/\varepsilon'_{rCNT}$ it is possible, using (17)-(20) to derive expression (13) for the imaginary part of the CNT permittivity based on the measured RX power, microwave Q factors of the CRA and S parameters of the RX and TX antennas.

## APPENDIX C. MEASURED |S11| PARAMETERS OF THE TX DIPOLE ANTENNAS

The measured $|S_{11}|$ data demonstrate that the reflected power from the CNT-loaded and unloaded CRAs is the same in the resonance transmission band.

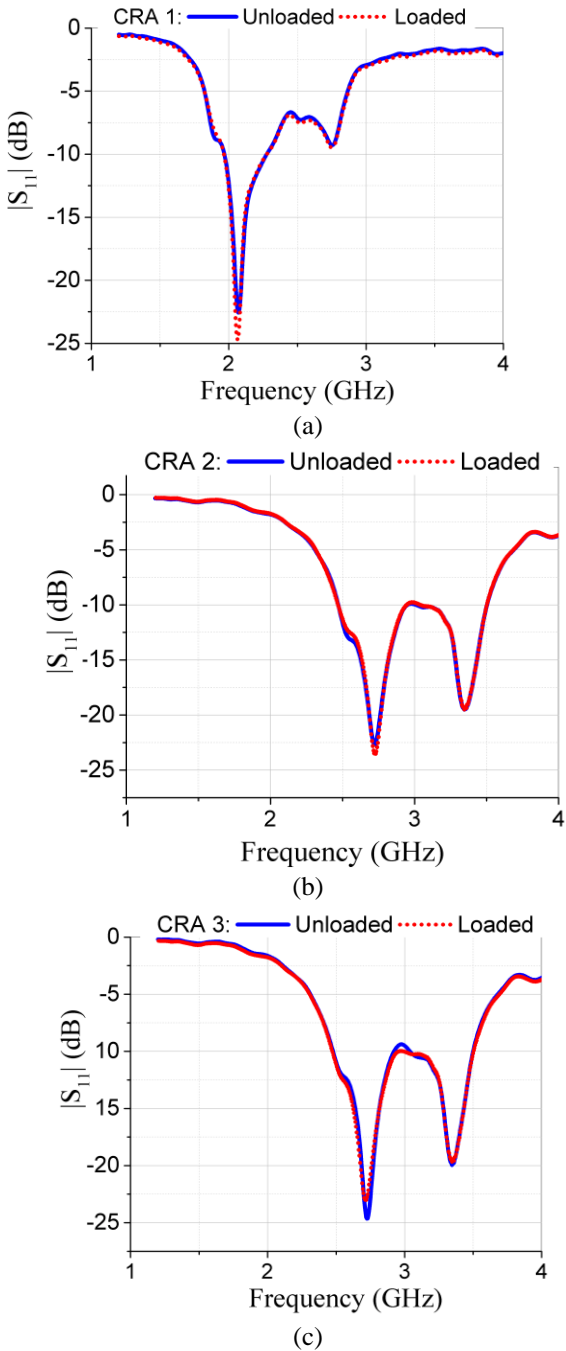

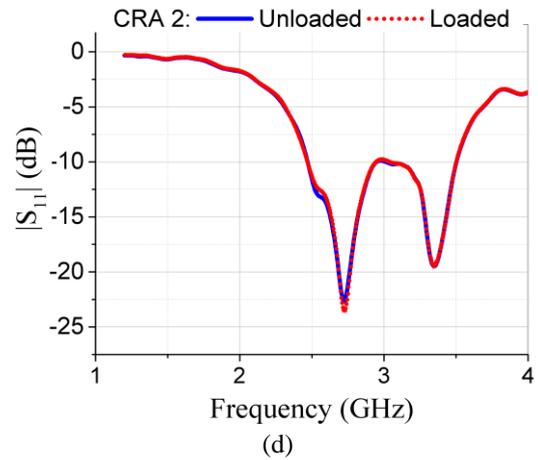

(d)

FIG. 6. (a)-(d) Reflection parameter $|S_{11}|$ for the CRAs 1-4.

## APPENDIX D. EM FIELD ENHANCEMENT INSIDE THE CRA

EM field enhancement inside the CRA gap is shown, as an example, for the CRA 2 using full-wave simulation data (obtained with FEKO solver, www.feko.info).

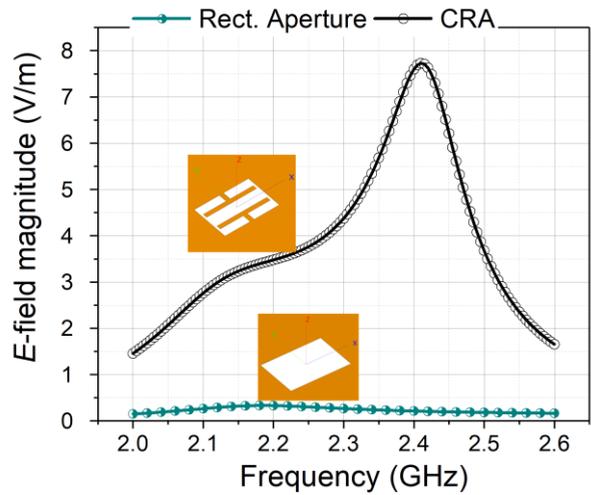

FIG. 7. E-field enhancement inside the CRA as compared to the rectangular aperture. Both apertures are fed with the same dipole antenna with port excitation power -25dBm.

Simulations presented in FIG.7 show that the E-field is enhanced mode than 36 times (31dB) inside the CRA as compared to the counterpart rectangular aperture (26mmx18mm).